\documentclass[useAMS,usenatbib]{mn2e}
\usepackage{graphicx}
\usepackage{natbib}  
\usepackage{epstopdf}
\newcommand\apj{{ApJ}}%   
\newcommand\apjl{{ApJ}}%   
\newcommand\aap{{A\&A}}%   
\newcommand\aaps{{A\&AS}}%   
\newcommand\mnras{{MNRAS}}%   
\def\simgt{\lower.5ex\hbox{$\; \buildrel > \over \sim \;$}}
\def\simlt{\lower.5ex\hbox{$\; \buildrel < \over \sim \;$}}
\newcommand\tbce{T$_{\rm bce}$}

\newcommand\alfa{$\alpha$}

\newcommand{\msun}{\ensuremath{\, {M}_\odot}}
\newcommand{\Msun}{\ensuremath{\, {M}_\odot}}
\newcommand{\ocen}{$\omega$~Cen}

\def\he3{$^3$He}
\title[Lithium production in low metallicity AGB stars]{The role of lithium production in massive AGB  
and super--AGB stars for the understanding of multiple populations in Globular Clusters
}
\author[P. Ventura, F. D'Antona]
{P. Ventura$^{1}$ \& F. D'Antona$^{1,2}$ \thanks{E-mail: ventura, dantona @oa-roma.inaf.it }\\
$^{1}$ INAF, Osservatorio Astronomico di Roma, Via Frascati 33, 00040 Monteporzio Catone (Roma), Italy.\\
$^{2}$ INAF, IASF--Roma, via Fosso del Cavaliere 100, I-00133 Roma, Italy\\
}
\begin{document}
\date{Accepted . Received ; in original form }

\pagerange{\pageref{firstpage}--\pageref{lastpage}} \pubyear{2006}

\maketitle

\label{firstpage}

\begin{abstract}
Lithium is made up in the envelopes of massive Asymptotic Giant Branch (AGB) stars through the
process of Hot Bottom Burning. In Globular Clusters, this processing is one possible source of
the hot-CNO burning whose nuclear products are then ejected into the intracluster medium 
and take part in the formation of a second stellar generation, explaining the peculiar distribution
of chemical elements among the cluster stars.
We discuss the lithium yields from AGB stars in the mass range 3-- $\sim$6.3\msun, and from super--AGB stars
of masses in the range $6.5$--9\msun\ for metallicity Z=10$^{-3}$. 
The qualitative behaviour of these yields is discussed in
terms of the physical structure of the different masses. Although many uncertainties affect 
the other yields of these stars (e.g. O, Na and Mg), even larger uncertainties affect
the lithium yield, as it depends dramatically on the adopted description of mass loss.
When we adopt our standard mass loss formulation, very large yields are obtained especially for the
super-AGB stars, and we discuss their possible role on the lithium 
abundance of second generation stars in globular clusters.
\end{abstract}


\section{Introduction}
\label{sec:intro}
In highly evolved giants, the pristine lithium
has been already well depleted from the stellar atmosphere due to convective dilution with 
the internal layers where it has been burned. If lithium is observed, its abundance 
is ascribed to temporary production due to the
action of the \cite{cameronfowler1971} mechanism: $^7$Be is produced by fusion of $^3$He with
$^4$He, and rapidly transported to stellar regions where it can be converted into $^7$Li
by electron-capture. Envelope models of asymptotic giant branch (AGB) stars 
\citep{scalo1975}, in which the temperature at the bottom of their hot bottom convective envelopes  
(\tbce) reaches the hydrogen burning layers, show that 
the $^3$He($\alpha,\gamma)^7$Be can act (at \tbce$\sim$40MK). These models were able to explain
the high lithium abundances found in some luminous red giants, and the process took the
name of Hot Bottom Burning (HBB). 
A different, slow mixing process named ``cool bottom processing" \citep{nollett2003} 
is instead at the basis of the lithium abundances seen in lower luminosity red giants 
\citep[e.g.][]{wasserburg1995,sackmann1999}. The physical reasons of this
process are not well studied, while the nucleosynthesis in HBB is simply based on 
time--dependent mixing in plain convection zones.
Apart from being a key process for the lithium production, HBB is
the key mechanism necessary to process to nitrogen the carbon present 
in the envelopes of carbon stars, that becomes active at slightly larger \tbce$\sim$65MK. 
Observationally, luminous AGB stars lose their carbon star 
character when they are lithium rich, although a small luminosity interval may be characterized
by stars that are both lithium and carbon rich \citep{ventura1999}, as in fact occurs in
the most luminous fraction of the J stars  \citep{abia1991,smith1995}.
In the most massive AGB stars, and especially in low metallicity
AGB stars, \tbce\ becomes larger than $\sim$80MK, and  also the ON chain of the CNO cycle becomes 
active. In these envelopes, oxygen is cycled to nitrogen, and its abundance can 
be dramatically reduced. Thus some models attribute the existence of 
anomalous stars in globular clusters (GCs), 
having low oxygen and high sodium, to the formation of a second stellar generation (SG) including matter 
processed by HBB \citep[e.g.][]{ventura2001}. Other models attribute the formation of the SG
to the ejecta of fast rotating massive stars \citep[FRMS, see e.g.][]{decressin2007a}, or even to
pollution from gas expelled during highly non-conservative evolution of massive 
binaries \citep{demink2009}, although this 
latter model in particular can not explain the very high fraction of SG stars present in
most of the GCs so far examined \citep{carretta2009a, carretta2009b}.
The lithium yield from AGB stars of different mass may contribute to understand the role (if any) of these
stars in the formation of the SG in GCs. 
It is commonly believed that the polluting matter must be diluted with pristine matter
to explain the abundance patterns, such as the Na--O anticorrelation (Prantzos \& Carbonnel 2005,
D'Antona \& Ventura 2007). If the progenitors of the SG stars are 
massive stars, for sure they do not have any lithium left, and the
lithium in the SG must be due to the mixing with pristine gas. If instead the progenitors are massive
AGB stars, they may have a non negligible lithium yield, that must be taken into account in the
explanation of the SG abundances.

%In this work we first summarize the observational evidence concerning the lithium spread in
%globular cluster stars, that appears to be larger than the halo stars spread in the same
%range of iron content, a possible indication of the lithium dilution in matter forming the SG stars
%(Section 2). Then we present the results of full computations of AGB and super--AGB stars evolution, and the 
%state of the art average lithium abundances in the envelopes of massive AGB stars of
%metallicity in the range spanned by galactic GCs (Section 3 and 4) and derive
%their implication for models of SG formation in GCs (Section 6).
%%
%%We study the dependence of 
%%the lithium yield on the initial mass and on  the assumptions made for the mass loss rate, and
%%provide physical explanations of the models' lithium trends. 
%%Although the lithium rich phase lasts for a short time, due to 
%%the very high \tbce's, the total yield in super--AGB stars becomes much larger than in the massive
%%AGB stars, so we discuss the implication of these results for 

\begin{figure}
\includegraphics[width=7cm]{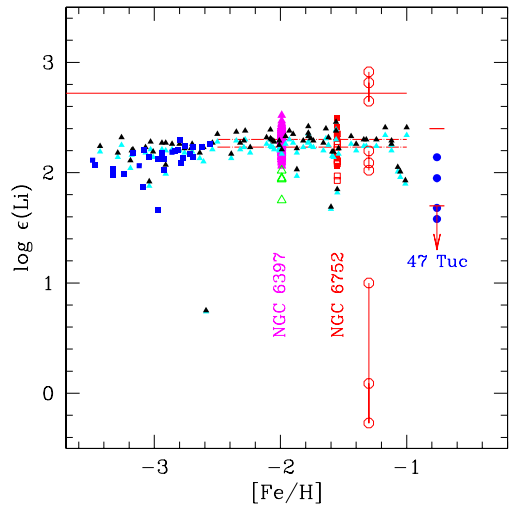}
\caption{Lithium abundances as a function of [Fe/H] in halo stars and in scarcely evolved 
stars in three GCs. Halo data are from Melenzez et al. (2009), represented as black triangles (LTE
models) or grey (cyan) triangles (non LTE models). (Blue) full squares are from Sbordone et al. (2009),
analyzed by 3D non LTE models. The two dashed horizontal lines represent an eye fit of the Melendez
LTE (upper line) and non LTE model data (lower line) in the range of the GC abundances. Data for
NGC~6397 are from Lind et al. (2009). The three open triangles are values for three
subgiants, and may not represent the turnoff abundances in this cluster. Data for NGC~6752 are 
from Pasquini et al. (2005), plotted as open or full squares according to the two different temperature
scales used in this work. The full circles are the data for 47 Tuc by Bonifacio et al. (2007). The limits
of the lithium range in the 50 stars recently examined by D'Orazi et al. (2009) are also given.
Open circles represent the average abundances in the ejecta of 
models of 4, 5 an 6\msun (see Table 1) for three different mass loss rate formulations.}
      \label{f1}%
\end{figure}

\section{The problems posed by observations}
Figure \ref{f1} shows a compact summary of what we know about lithium 
abundances in the halo and in GCs in the plane log$\epsilon$(Li) versus 
[Fe/H] \footnote{the standard notations, i.e. log$\epsilon$(Li)=log(Li/H)+12 and
[Fe/H]=log(Fe/H)-log(Fe/H)$_{\odot}$, were adopted}. The halo stars are plotted as triangles, from
Melendez et al. (2009). We show both the LTE and non LTE abundances (these latter as grey triangles).
The data for three clusters are added, at their [Fe/H] content, taken from \cite{carretta2009c} scale. 
%For NGC 6397 we plot the data by \cite{lind2009}, for NGC 6752 the data by \cite{pasquini2005}, and for 47~Tuc 
%the data by \cite{bonifacio2007}. For the latter cluster,we also show the range of abundances obtained 
%in the very recent investigation by \cite{dorazi2009}. 
The references for the clusters data are in the Figure label. Notice that
the three open triangles of NGC~6397, at much lower $\epsilon$(Li) than the other points, 
refer to subgiants, in which Lithium is reduced by mixing. Although the data analysis 
is not homogeneous among the different samples, the figure shows interesting trends.
The lithium spread in the field stars in the range of metallicities of the 
clusters NGC~6397 and NGC~6752 is very small around a plateau value 
log$\epsilon$(Li)$\sim$2.2 for the non LTE abundances, and slightly larger for the LTE
determinations. In fact the triangles at log$\epsilon$(Li)$<2$\ are lower mass stars for which depletion
is expected \citep{melendez2009}. The WMAP -- big bang nucleosynthesis 
``standard" abundance, log$\epsilon$(Li)=2.72 \citep{cyburt2009} is much larger than the plateau abundance.
The lithium spread in the clusters appears a bit larger, although \cite{lind2009} 
point out that in NGC~6397 it is consistent with the observational error.
We should expect a larger lithium spread among GC stars if there are SG stars, even if
the pollutors' gas (AGB or massive stars envelopes) has been diluted 
with pristine gas \citep{decressin2007b,prantzos2007}. The dilution is very plausible
if there is a direct correlation 
between lithium and sodium abundances, as convincingly shown in NGC~6752
\citep{pasquini2005}. A similar correlation in NGC~6397 
is based only on the high sodium abundance of the three subgiants 
plotted as open triangles \citep{lind2009}, and is not
convincing among the stars of 47~Tuc \citep{bonifacio2007} that may suffer
from normal depletion mechanisms due to their larger iron content \citep{dorazi2009}. 
In addition, according to \cite{pasquini2008}, two stars in NGC~6397 differ by $\sim$0.6dex
in oxygen, but have ``normal" log$\epsilon$(Li)$\sim$2.3: this is certainly not easily compatible with 
a simple dilution model, and may require that the pollutors are also important lithium producers.
In fact, if the AGB pollutors produce enough lithium, a dilution model must take it into account.
Notice that the dilution model is not so straightforward as we may think a priori: it 
will include a fraction \alfa\ of
matter with pristine Li, plus a fraction (1-\alfa) having the Li of the ejecta (so, either the abundance
of the AGB ejecta in the AGB mass range involved in the SG formation, or zero Li for the FRMS model).
The dilution required to explain a given range in observed Li is different 
if we assign to the pristine Li the value log$\epsilon$(Li)=2.72 (see above), or
the atmospheric pop.II value ($\sim$2.36), or some intermediate value. In addition, if we 
are assuming that the uniform surface abundance of Li in pop.II is due to a depletion
mechanism, also the abundance resulting from the dilution model must be decreased to take into
account a similar depletion factor.

A different interesting problem is posed by the GCs in which a ``blue" 
main sequence (MS) has been revealed
from precise HST photometry, namely \ocen \citep{bedin2004} and NGC~2808. 
The blue MS can only be interpreted as a very high helium (mass fraction Y$\sim$0.38) 
MS \citep{norris, piotto2005}. 
Actually, in NGC~2808 three MS well separated each other in color are present  
\citep{piotto2007}, corresponding to three main helium content values, 
and in agreement with the predictions made from the distribution of stars in the
very extended and multimodal horizontal branch \citep{dc2004,dantona2005}.
As the blue MSs are well detached in color, 
all their stars must have helium content in a very strict range of values ---practically 
a unique value of helium. This poses a problem for the FRMS model, in which the stars are formed
in the discs ejected by the progenitors. In this model, the SG helium abundance will reflect the 
abundance in the disc, that will vary from one massive star to another, or even will be linked
to the evolutionary stage at which the stellar envelope matter is lost \citep{decressin2007b}. 
Consequently, 
stars born out of such discs will show a spread in helium abundances, that will result in a
broadened MS rather than in a well detached blue MS \citep{renzini2008}. 
On the contrary, \cite{pumo2008} noticed
that the helium abundances of super--AGB stars envelopes are within the small range 0.36$<$Y$<$0.38
\citep{siess2007}, and \cite{dercole2008} have shown that a full chemo--hydrodynamical
model of the cluster can provide a reasonable interpretation 
of the three MSs of NGC~2808, {\it provided that
the blue MS is formed directly by matter ejected from the super--AGB range, undiluted with 
pristine gas}. In the future, spectroscopic observations of the blue MS in \ocen\ and NGC~2808
will provide a falsification of this hypothesis, e.g. by means of the oxygen and sodium abundance
revealed. In particular lithium can be an important test too, as it could provide an
independent calibration of the mass loss rate in the super--AGB phase.
Already some observations of lithium at the turnoff of 
\ocen\ are available (Bonifacio et al. 2009, in preparation), but it is not clear whether they contain
stars belonging to the blue MS. 
Our computations thus have two aims: 1) to compute the average lithium abundance 
in the envelopes of different evolving AGB masses, and show its dependence on the
mass loss formulation; 2) to provide the first computation of the super--AGB stars lithium yields, when 
using the ``standard" mass loss rate formulation adopted for the massive AGB stars.
The results presented here can be applied to model the lithium abundance in the SG of 
different GCs.

\section{The models}
In this work we show the lithium evolution obtained in our recent AGB models,
discussed in the series of papers by \cite{ventura2005a, ventura2005b}, and 
whose yields are summarized in Table 2 of \cite{ventura2009}. 
We deal with a metallicity of Z=10$^{-3}$\ in all the models described here.
The numerical structure of the evolutionary code  ATON is described in details in \cite{ventura1998}.
We adopt the latest opacities by \cite{ferguson2005} at temperatures 
lower than 10000 K and the OPAL opacities in the version documented by \cite{iglesias1996}. 
The mixture adopted is $\alpha$--enhanced, with $[\alpha$/Fe$]=0.4$ \citep{GS98}.
The conductive opacities are taken from Poteckhin (2006, 
see the web page www.ioffe.rssi.ru/astro/conduct/), 
and are harmonically added to the radiative opacities.
Details on the rest of the physical inputs (equation of state, convection, overshooting, 
nuclear reaction rates) can be found in \cite{ventura2009}.
%Tables of the equation of state are generated in the (gas) pressure-temperature
%plane, according to the latest release of the OPAL EOS (2005), overwritten in
%the pressure ionization regime by the EOS by Saumon, Chabrier \& Van Horn (1995),
%and extended to the high-density, high-temperature domain according to the
%treatment by Stoltzmann \& Bl\"ocker (2000). 
%Convection was modelled according to the FST prescription.
Time--dependent
mixing of chemicals within convective zones has been treated as a
diffusive process, following the approach by Cloutman \& Eoll (1976), solving 
for each chemical species the diffusive-like equation:
$$
{dX_i\over dt}=\big( {\partial X_i\over \partial t} \big)_{nucl}+
{\partial \over \partial m_r} \big[ (4\pi r^2\rho )^2D{\partial X_i \over \partial m_r} \big]
\eqno{(1)}
$$
where D is the diffusion coefficient, for which, given the convective velocity
$v$ and the scale of mixing $l$, a local approximation ($D\sim {1\over 3}vl$) 
is adopted. The borders of the convective regions are fixed according to the
Schwarzschild criterion. 
%We considered extra-mixing from all the formal
%convective boundaries up to the beginning of the AGB phase: convective 
%velocities are assumed to decay exponentially with an e-folding distance 
%described by the free-parameter $\zeta$, that was set to $\zeta=0.02$,
%according to the calibration provided in Ventura et al. (1998), where the
%interested reader can also find a complete discussion regarding the variation 
%of the convective velocities in the proximities of the convective borders.
In our ``standard" models, mass loss is described according to the Bl\"ocker (1995) formulation, that
extends Reimers' recipe to describe the steep 
increase of mass loss with luminosity as the stars climb the AGB on the
HR diagram. The full expression, for Mira periods exceeding 100d, is
$$
\dot M=4.83 \times 10^{-22} \eta_R M^{-3.1}L^{3.7}R
\eqno{(2)}
$$
where $\eta_R$ is the free parameter entering the Reimers' (1977) prescription.
We use $\eta_R=0.02$, according to the calibration based on
the luminosity function of lithium rich stars in the Magellanic Clouds 
given in Ventura et al.(2000).
In this work we also consider an extreme value of $\eta_R=0.1$, and models adopting the
Vassiliadis \& Wood (1993) mass loss rate.

For core masses above $\sim$1.05\msun ---for these models, above initial mass 6.3\Msun--- 
carbon ignites in the C--O core, in conditions of semi--degeneracy
\citep{ritossa1999,siess2007} and the star acquires a degenerate O--Ne core that can evolve through
thermal pulses like an AGB star. 
We follow the lithium phase in the super--AGB models of masses M$\geq$6.5\msun. 
The details of these models are presented elsewhere (Ventura 2009, in preparation). 

\begin{figure}
\includegraphics[width=7cm]{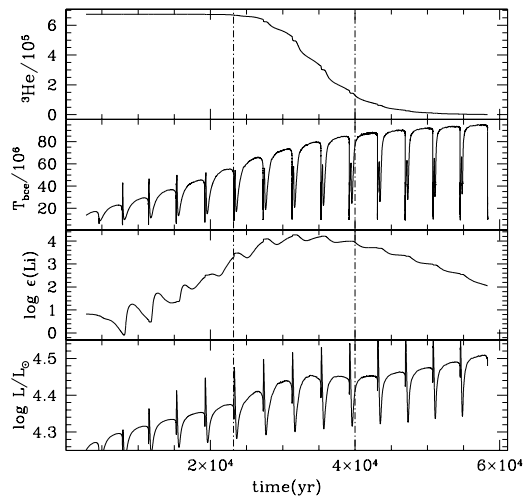}
\caption{From bottom to top panel we plot the luminosity, surface lithium,
 HBB temperature and $^3$He surface content along the AGB evolution of the 5\Msun\ with standard
 $\eta_R$=0.02 mass loss rate. The two vertical lines delimit the time of maximum lithium production,
when the surface lithium abundance roughly exceeds log$\epsilon$(Li)=3.}
      \label{f2}%
\end{figure}

\begin{table}
\caption{Lithium average abundances in the ejecta for Z=10$^{-3}$}             
\label{mloss}      
\centering          
\begin{tabular}{c c c c }     % 4 columns 
\hline\hline       
$M/M_{\odot}$ & $\eta_R$=0.02 & $\eta_R$=0.1 & VW rate   \\ 
\hline            
4.00  &  2.197  &   2.816   &   0.089 \\
5.00  &  2.089  &   2.647   &  --0.27 \\
6.00  &  2.021  &   2.914   &   1.040 \\
\hline
\end{tabular}
\begin{tabular}{c c c c}     % 4 columns 
\hline\hline       
$M/M_{\odot}$ & log $\epsilon$(Li) & M$_{core}$/M$_\odot$ & core  \\ 
\hline         
3.00  &  2.771 & 0.76 & CO \\
3.50  &  2.438 & 0.80 & CO  \\
4.00  &  2.197 & 0.83 & CO \\
4.50  &  1.998 & 0.86 & CO  \\
5.00  &  2.089 & 0.89 & CO  \\
5.50  &  1.757 & 0.94 & CO  \\
6.00  &  2.021 & 1.00 & CO \\
6.30  &  2.078 & 1.03 & CO \\
6.50  &  2.350 & 1.09 & O-Ne\\
7.00  &  1.950 & 1.20 & O-Ne\\
7.50  &  2.910 &1.27 &  O-Ne\\
8.00  &  4.480 & 1.36 &  O-Ne\\
\hline\hline
\end{tabular}
\end{table}

\section{Results}
We show in Figure \ref{f2} the time evolution of luminosity, surface lithium abundance, temperature
at the bottom of the convective envelope (\tbce) and $^3$He surface abundance along the evolution of the
standard ($\eta_R$=0.02) model of 5\msun. The time variable is set to zero at the end of the
core helium burning phase. We see that the HBB phase begins when \tbce$\sim 40\times 10^6$K,
but the lithium-rich phase, when log$\epsilon$(Li) exceeds 3, occurs when \tbce$\simgt 60\times 10^6$K, 
and it ends $\sim 2\times 10^4$yr after, due to the exhaustion of the $^3$He in the envelope.

\begin{figure}
\includegraphics[width=7cm]{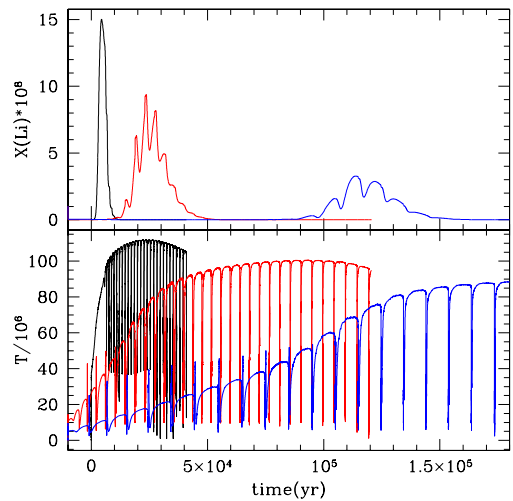}
\caption{Temperature at the bottom of the convective layer (bottom) and lithium surface abundance
(top) as a function of the time for the masses 6, 5 and 4\msun, from left to right. Time is computed
from the beginning of the AGB phase, when the H--shell burning is reignited.}
      \label{f3}%
\end{figure}

\begin{figure}
\includegraphics[width=7cm]{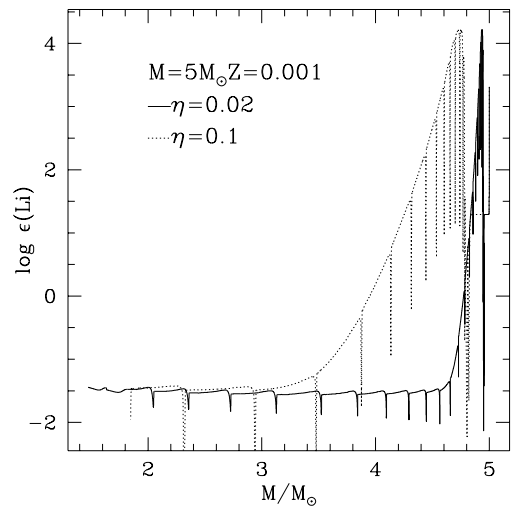}
\caption{The surface lithium abundance as a function of the stellar mass, during the evolution
with mass loss starting at 5\Msun. The standard case $\eta_R$=0.02 is compared to the larger
mass loss rate case.}
       \label{f4}%
\end{figure}

Figure \ref{f3} displays the lithium production and \tbce\ in the envelopes of stars of 6, 5 and 4\msun. 
We adopt a linear scale for the mass fraction of lithium (X$_{\rm Li}$) that shows more
clearly the timescale of production and destruction. The larger masses achieve a larger 
peak of abundance, but their production timescale is shorter\footnote{Notice
that also the {\it total duration} of the thermal pulse phase is shorter, for larger
masses, due to the faster mass loss.}. This is due to the larger
\tbce, that causes a faster consumption of the $^3$He buffer. 
The lithium yield will depend on the lithium abundance reached in the envelope,
on the duration and on the mass lost during the phase of lithium production. 
In Table 1 we show the average abundance in the whole ejected envelopes (the yield
is obtained by multiplying the abundance to the total envelope mass, equal to 
the difference between the initial mass and the core mass listed in the third column). 
For 4, 5 and 6\msun\ we also show
the result for the three different mass loss rate formulations.
In the extreme case of mass loss ($\eta_R$=0.1) shown in the Table, we see that average abundances 
0.6--0.7dex larger than the standard ones are reached, while the abundances obtained by 
adopting \cite{vw1993} mass loss formulation are negligible. The comparison of the lithium production
during the 5\Msun\ standard and $\eta_R$=0.1 models is shown in Figure \ref{f4}, the abundances for these
models are shown as open circles in Figure \ref{f1}.
The standard models ($\eta_R$=0.02) show a non monotonic behaviour: the yield from the 4\msun 
is larger than at 5\msun,
in spite of the lower peak abundance reached in the envelope, due to the longer duration of
the lithium rich phase. However, the 6\msun\ yield is also larger than the 5\msun\ yield, thanks both to
the larger abundances and to the larger luminosity of the star, that favours a larger mass loss rate.
The evolution of the 6\msun\ is also characterized by the fact that the HBB and the lithium 
production begin before the thermal pulse phase. 
\begin{figure}
\includegraphics[width=7cm]{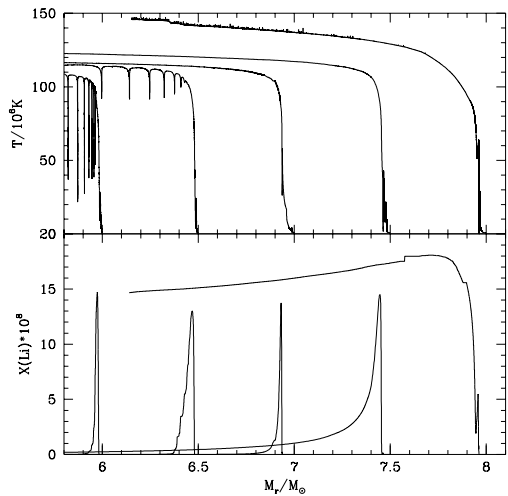}
\caption{Temperature at the bottom of the convective layer (top) and lithium surface abundance
(bottom) as a function of the total mass of the mass losing stars of 6, 6.5, 7, 7.5 and 8\msun.}
      \label{f5}%
\end{figure}
Table 1 also includes the standard results with mass spacings of 0.5\msun. 
The behaviour of the lithium average abundance is not perfectly
monotonic, but there is a general trend (decreasing abundances for increasing initial mass) 
that is inverted as soon as we deal with the super--AGB models.
The physical conditions in super--AGB stars are in fact more extreme: their
luminosity is larger, so that we have to deal with larger mass loss rates, and they achieve a larger \tbce, so that 
more extreme HBB nucleosynthesis is expected. 
HBB begins very early during the star evolution. Increasing the mass, the entire lithium production
and destruction phase occurs before the onset of thermal pulses, as we show in Figure \ref{f5}. Given the chosen
mass loss law, the 8\Msun\ evolution actually may not even reach the thermal pulse phase. The yield of the
8\msun\ becomes enormous. Of course, a different mass loss law may dramatically decrease
the lithium average abundances.
%Figure \ref{f6} shows the lithium average abundance in the ejecta as a funtion of the total initial mass
%for the whole mass range from 3 to 8\msun. Apart from the discontinuities, the trend is clear:
%first the lithium production decreases when increasing the mass, as the lithium rich phase becomes shorter, then
%the much higher mass loss rate, and the increased peak lithium achieved in the envelope 
%are such that the lithium production increases with the mass.

\section{Conclusions. Lithium in the second generation of galactic globular clusters}

Our results show that the lithium yield from massive AGB stars is dramatically dependent on the adopted
mass loss formulation. If we take our ``standard" results in Table 1 at face value, ignoring the
big question mark on mass loss, the yields can be used to predict the lithium expected in the
SG, if the SG is a result of star formation from AGB ejecta diluted with pristine gas. The abundances
will depend mainly on the mass range of the AGB progenitors: if the ejecta of 
masses in the range 4.5 -- 6\msun\ are involved, their abundance is log$\epsilon$(Li)$\sim$2, 
0.7dex smaller than the Big Bang abundance. In order to explain the abundances observed in NGC~6397 
or in NGC~6752, a dilution model including the ejecta of these AGB stars
will require only a slightly smaller percentage of pristine matter, with respect 
to a model including the lithium free FRMS, and we can not discriminate between the two models.  
The situation is different for the cases in which super--AGB ejecta may be involved. In fact
the main result of these computations is that the lithium
yield becomes very large in the super--AGB evolution: it may become much larger
than the population I abundance for progenitor masses of $\sim$8\msun. This star
has an O--Ne core of 1.36\msun, so it is close to
the mass limit of the super--AGB range. Of course, this very high lithium production is linked to our
own modelling of the physics of massive AGB stars and super--AGB stars, and in particular 
to the choice of the mass loss law. More generally, these models predict that the stars belonging 
to the blue main sequence of $\omega$ Cen and NGC~2808 can be richer in lithium than the other SG stars in GCs, if they were born
directly from super--AGB ejecta, as we propose to explain their high and uniform helium abundance.
If this prediction will be demonstrated to be incorrect, the mass loss rates 
from super--AGB stars (and massive AGB stars as well) must be revised downwards. 
As a consequence, the lithium production from massive AGB stars  will not
be significant, and both dilution models, either with AGB stars or with FRMS matter, will not provide 
different informations for lithium. We then urge that a strong observational effort is made to observe 
spectroscopically the blue MS stars.
%\begin{figure}
%\includegraphics[width=7cm]{litio7.eps}
%\caption{}
%      \label{f6}%
%\end{figure}

\section{Acknowledgments} 
This work has been supported through PRIN MIUR 2007 
``Multiple stellar populations in globular clusters: census, characterization and
origin" (prot. n. 20075TP5K9). We thank V. D'Orazio, J. Melendez, D. Romano and L. Sbordone, 
for information, sharing data and discussions.

\label{lastpage}

\end{document}